\documentclass[aps,floats,prb,showpacs,twocolumn ]{revtex4}

\usepackage{amsfonts,amsmath} \usepackage{bm} \usepackage{dcolumn}
\usepackage{epsfig} \usepackage{latexsym}

\begin{document}

\title{Supersymmetric field theory of local light diffusion in semi-infinite media}

\author{Chushun Tian}

\affiliation{Institut f{\"u}r Theoretische Physik, Universit{\"a}t
zu K{\"o}ln, K{\"o}ln, 50937, Germany }

\date{\today}


\begin{abstract}

{\rm A supersymmetric field theory of light diffusion in
semi-infinite disordered media is presented. With the help of this
technique we justify--at the perturbative level--the local light
diffusion proposed by Tiggelen, Lagendijk, and Wiersma [Phys. Rev.
Lett. \textbf{84}, 4333 (2000)], and show that the coherent
backscattering line shape of medium bar displays a crossover from
two-dimensional weak
to quasi-one-dimensional strong localization.}

\end{abstract}

\maketitle

\section{Introduction}
\label{intro}

The Anderson localization of light has been one of the most
fascinating phenomena in condensed matter physics since the
mid-eighties \cite{John84,Anderson85}. Like electron systems this
phenomenon finds its origin in coherent multiple scattering which
slows down diffusion of photons and eventually brings them to stop.
Parallel to studies of disordered conductors the subject in this
field ranges from light localization near or far below the mobility
edge in bulk (infinite) systems \cite{John84,Kroha93} (where
low-energy photon motion enjoys the translational symmetry) to their
detection such as transmission measurements in the slab geometry
(e.g., Refs.~\onlinecite{Chabanov01,Zhang03,Maret06}).

A unique subject of localization in optical (and other classical
wave) systems is the enhanced coherent backscattering (CBS)
phenomenon \cite{Golubentsev,Niuwenhuizen}. In this subject the
issue of semi-infinite geometry is heavily addressed because the CBS
line shape is responsible for by optical paths near the
vacuum-medium interface. Although it is well known that
in the weak disorder region, i.e., $l\gg \lambda$ ($l$ the mean free
path and $\lambda$ the wavelength) incident photons enjoy diffusion
as in bulk media \cite{Golubentsev,Niuwenhuizen}, in the strong
disorder region $l\lesssim\lambda$ the role played by the leakage at
the interface has been of long term interests \cite{Berkovits87}
and, still, remains in the central position of CBS studies,
particularly to forecast or observe the CBS line shape
\cite{Edrei90,Lagendijk97,Lagendijk00,Zhang02}.

Pressingly, in the latter region strong localization emerges in the
bulk and a new scale namely the localization length $\xi$ appears.
On the experimental side, there has been increasing evidence
indicating that inside the boundary layer of thickness $\gtrsim \xi$
the photon leakage at the vacuum-medium interface strongly
interplays with strong localization
\cite{Zhang03,Maret06,Lagendijk97,Zhang07}. On the theoretical side,
some time ago exact solution of semi-infinite one-dimensional
disordered chains shed the light on the existence of so-called
radiative localization states in the boundary layer
\cite{Chernyak92}, which lead to anomalous slowing decay of
reflected (backscattered) incident light pulses
\cite{Chernyak92,Zhang87}.

Recently, in an insightful theoretical work \cite{Lagendijk00} it
was realized that (in three-dimensional disordered media) inside the
boundary layer the translational symmetry of low-energy
(hydrodynamic) photon motion is strongly destroyed resulting in the
so-called ``local diffusion''. Remarkably, constructive wave
interference renders the static diffusion coefficient depending on
the distance from the interface. Consequently, the local diffusion
was found to lead to a rounded CBS line shape resembling that
observed experimentally \cite{Lagendijk97} and, thus, might overcome
the conceptual difficulty of earlier theoretical proposal
\cite{Berkovits87}. Surprisingly, the dynamic generalization of the
local diffusion equation \cite{Skipetrov04,Skipetrov06} provides an
explanation of some key phenomena observed in quasi-one-dimensional
microwave experiment \cite{Zhang03}, and well captures anomalous
slowing decay of reflected incident pluses in both
quasi-one-dimensional \cite{Zhang87,Skipetrov04,Beenakker00} and
three-dimensional disordered media \cite{Skipetrov06}. Moreover,
such novel prediction--the position dependence of diffusion
coefficient--seems to have been confirmed by a very recent
experiment \cite{Zhang07}.

Despite of these progress theoretical investigations on local
diffusion are, however, restricted on the self-consistent
diagrammatical method \cite{Lagendijk00,Skipetrov04,Skipetrov06}.
Thus, an intellectual challenge is to seek for the genuine
microscopic origin underlying this novel concept. This is, indeed,
the purpose of this work. The last few decades have witnessed
spectacular success of applications of supersymmetric field theory
to various disordered systems in the absence of interactions (Such
condition is perfectly satisfied by optical systems.)
\cite{Efetov97}. Among them there are few exact nonperturbative
results for quasi-one-dimensional disordered wires such as
density-density correlation function (in the infinite geometry)
\cite{Efetov83} and transmission statistics \cite{Lamakraft}. They
allow one to make important insight on the strong localization. Most
importantly, for periodic disordered media by using the
supersymmetric field-theoretic method a local light diffusion
equation, similar to that proposed in Ref.~\onlinecite{Lagendijk00},
recently has been derived at the microscopic level \cite{Tian07}. In
view of these it is natural and inevitably necessary to proceed
along the same line to explore the concept of local light diffusion
and its effects for more general--fully disordered--media, which
differ drastically from the former \cite{Tian07} from both physical
and technical view.

The main results of this paper are as follows. (i) We present a
field-theoretic proof showing that, in contrary to the conjecture of
Ref.~\onlinecite{Berkovits87}, no scaling behavior exists inside a
layer of thickness $\sim l$ extrapolating into the vacuum. (ii) We
justify--at the perturbative level--the local diffusion equation
proposed in Refs.~\onlinecite{Lagendijk00,Skipetrov04}. (iii) We
analyze signatures of the static local diffusion in the CBS line
shape. It should be stressed that in this paper the suspersymmetric
field theory is treated perturbatively, and the nonperturbative
treatise will be reported in the forthcoming paper.

The rest of this paper is organized as follows. In the next section
we produce nonlinear supersymmetric $\sigma$ model in the context of
optical systems. Most importantly, we derive the boundary constraint
satisfied by the supersymmetric matrix field. The supersymmetric
field theory is then applied to the two-dimensional medium bar.
Sec.~\ref{renormal} is devoted to exploring states residing deeply
inside the semi-infinite medium bar (namely far away from the
interface) by investigating renormalization effects of infinite
medium bar. In Sec.~\ref{WLsemiinfinite} weak localization in the
semi-infinite medium bar is studied, where the general dynamic
local diffusion equation is justified. The static limit of the local
diffusion equation is studied in Sec.~\ref{localdiffueq}. In
particular, the weak localization correction to the bare diffusion
constant is explicitly calculated, and its signatures in the CBS
line shape are analyzed. We conclude in Sec.~\ref{conclusion} and
give some technical details in
Appendix~\ref{DerivationEffGreen}-\ref{Izdiscussion}.

\section{Supersymmetric field-theoretic formalism}
\label{SUSY}


In this section a supersymmetric field-theoretic formalism is
presented for light scattering in semi-infinite disordered medium.

\subsection{Nonlinear $\sigma$ model}
\label{NSM}

We first show that as interactionless electron systems low-energy
photon motion in bulk disordered media is well described by the
nonlinear $\sigma$ model. The derivation is rather standard
\cite{Efetov97}. Here we only outline the scheme with an emphasis on
the main difference, while refer the reader to
Ref.~\onlinecite{Efetov97} for the details.

In the present work for simplicity the scalar wave will be
considered. The wave propagation in a bulk disordered medium is
described by the Helmholtz equation:
\begin{equation}
\left\{\nabla^2 + \Omega^2\left[1+\epsilon({\bf
r})\right]\right\}E({\bf r}) = j({\bf r}) \,,
\label{Helmholtz}
\end{equation}
where the field $E$ has the radiation frequency $\Omega$ (velocity
$c$ set to be unity), and $j({\bf r})$ is the source. Here the
fluctuating dielectric field $\epsilon({\bf r})$ has zero mean and
is distributed according to the Gaussian $\delta$-correlated law:
\begin{equation}
\Omega^4 \left\langle \epsilon({\bf r})\epsilon({\bf
r}')\right\rangle = \Delta \, \delta({\bf r}-{\bf r}') \,.
\label{dielectric}
\end{equation}

The Helmholtz equation resembles the Schr{\"o}dinger equation with
the Hamiltonian now read out as ${\hat H}=-\nabla^2 -\Omega^2 \,
\epsilon ({\bf r})$\,. As usual we may introduce the
retarded/advanced Green function $G^{R,A}_{\Omega^2}$ defined as
\begin{equation}
\{
 \Omega_\pm^2- {\hat H} \}\, G^{R,A}_{\Omega^2}({\bf r},{\bf
 r}')=\delta({\bf r} -{\bf r}') \,,
\label{Greenfunction}
\end{equation}
where $\Omega_\pm = \Omega \pm i0^+ $\,. The electric field and the
source are related via $E({\bf r})=\int d{\bf
r}'\,G^{R}_{\Omega^2}({\bf r},{\bf
 r}')j({\bf r}') $\,. We may further introduce the diffuson ${\cal Y}^{\rm
 {D}}$ and the cooperon ${\cal Y}^{\rm {C}}$ propagator defined as
\begin{eqnarray}
{\cal Y}^{\rm {D}} ({\bf r},{\bf r}';\omega) &  \equiv  &
\overline{G^R_{(\Omega+\omega^+/2)^2} ({\bf r},{\bf
 r}') \, G^A_{(\Omega-\omega^+/2)^2} ({\bf r}',{\bf r})}\,,  \nonumber\\
{\cal Y}^{\rm {C}} ({\bf r},{\bf r}';\omega) &  \equiv  &
\overline{G^R_{(\Omega+\omega^+/2)^2} ({\bf r},{\bf
 r}') \, G^A_{(\Omega-\omega^+/2)^2} ({\bf r},{\bf r}')}
 \label{DCdefinition}
\end{eqnarray}
with $\omega^+=\omega+i0^+$ and $\omega \ll \Omega$\,, where the
overline stands for the average over random dielectric field. These
two propagators
describe elegantly the light propagation over large scales.

The propagators above are represented in terms of superintegrals.
For this purpose we define a supervector field $\psi$\,:
\begin{eqnarray}
\psi=\left(
       \begin{array}{c}
         \psi^1 \\
         \psi^2 \\
       \end{array}
     \right)\,,
\psi^m = \frac{1}{\sqrt 2}\, \left(
           \begin{array}{c}
             \chi^{m*} \\
             \chi^m \\
             S^{m*} \\
             S^m \\
           \end{array}
         \right)\,, m=1,2
 \label{supervector}
\end{eqnarray}
with $S$'s ($\chi$'s) complex commuting (anticommunting) variables,
where the superscript $1$ ($2$) refers to retarded (advanced) Green
function, and its charge conjugate ${\bar \psi}\equiv \psi^\dagger
\Lambda$\,. Here $\Lambda$ is an $8\times 8$ supermatrix:
\begin{equation}
\Lambda = \left(
            \begin{array}{cc}
              1 & 0 \\
              0 & -1 \\
            \end{array}
          \right)^{\rm ar} \otimes \mathbf{1}^{\rm bf} \otimes \mathbf{1}^{\rm
          tr}\,.
 \label{lambda}
\end{equation}
Hereafter supermatrices are defined on the retarded/advanced (ar),
bosonic/fermonic (bf) and time-reversal (tr) sector. Then
\begin{eqnarray}
&  &  {\cal Y}^{\rm {D}} ({\bf r},{\bf r}';\omega) \nonumber\\
&  =  &  -4  \int d[\psi]\, \psi^1_\alpha({\bf r}){\bar
\psi}^1_\alpha({\bf r}')
\psi^2_\beta({\bf r}'){\bar \psi}^2_\beta({\bf r})\, \overline{e^{-\mathfrak{L}[\psi,{\bar \psi}]}}\,,  \nonumber\\
&  &  {\cal Y}^{\rm {C}} ({\bf r},{\bf r}';\omega) \nonumber\\
&  =  &  -4  \int d[\psi]\, \psi^1_\alpha({\bf r}){\bar
\psi}^1_\alpha({\bf r}') \psi^2_\beta({\bf r}){\bar
\psi}^2_\beta({\bf r}')\, \overline{ e^{-\mathfrak{L}[\psi,{\bar
\psi}]}}\,.
\label{DCsuper}
\end{eqnarray}
Here
\begin{equation}
\mathfrak{L}=i\int {\bar \psi}({\bf r}) \left[-{\cal H}_0-\Omega^2
\epsilon({\bf r})-\Omega\, \omega^+\Lambda \right]\psi({\bf r})
d{\bf r}\,.
 \label{Lagrange}
\end{equation}
with ${\cal H}_0=-\nabla^2-\Omega^2$\,, where the $\omega^2$ term is
omitted since $\omega \ll \Omega$\,. Performing the average we
arrive at
\begin{equation}
\overline{e^{-\mathfrak{L}[\psi,{\bar \psi}]}} = e^{-i\int {\bar
\psi} \left[-{\cal H}_0 - \Omega\, \omega^+\Lambda\right]\psi \,
d{\bf r}-\frac{\Delta }{2}\,\int \left({\bar \psi}\psi\right)^2
d{\bf r}} \,.
 \label{Lagrangeaverage}
\end{equation}

The quartic term is decoupled by the standard Hubbard-Stratonovich
transformation. Introduce an $8\times 8$ supermatrix field $Q({\bf
r})$ conjugate to $\frac{2}{\pi N(\Omega^2)}\, \psi({\bf r})\otimes
{\bar \psi}({\bf r})$\,. Here $N(\Omega^2)$ is related to the photon
density of states per unit volume $\nu(\Omega)$ by
$\nu(\Omega)=2\Omega N(\Omega^2)$\,.
Then,
\begin{eqnarray}
&  &  \exp\left[-\frac{\Delta}{2}\, \int \left({\bar
\psi}\psi\right)^2\, d{\bf r} \right] \label{HStransformation}  \\
&  =  & \!\!\int\!\! \exp\left[-\pi\Delta N(\Omega^2)\!\!\int \!\!
\left(\! {\bar \psi}Q\psi+\frac{\pi N(\Omega^2)}{4}\, Q^2 \!
\right)d{\bf r}\right]\!\! D[Q] \,. \nonumber
\end{eqnarray}
Substituting it into Eqs.~(\ref{DCsuper}) and
(\ref{Lagrangeaverage}) and integrating out the $\psi$ fields using
the Wick theorem, we obtain:
\begin{eqnarray}
&  &  {\cal Y}^{\rm {D,C}} ({\bf r},{\bf r}';\omega) =\left[\frac{\pi N(\Omega^2)}{4}\right]^2 \times \label{DC} \\
&  &  \langle{\rm str} [k(1+\Lambda)(1- \tau_3) Q({\bf r})
(1-\Lambda)(1\mp \tau_3) kQ({\bf r}')] \rangle \nonumber
\end{eqnarray}
with
\begin{equation}
k = \left(
            \begin{array}{cc}
              1 & 0 \\
              0 & -1 \\
            \end{array}
          \right)^{\rm bf} \otimes \mathbf{1}^{\rm ar} \otimes \mathbf{1}^{\rm tr}
 \label{lambda}
\end{equation}
and $\tau_k$ the Pauli matrices defined on the time-reversal sector.
In Eq.~(\ref{DC}) the following average is introduced:
\begin{equation}
\langle P[Q]\rangle = \int
D[Q]\, P[Q]\, e^{-F[Q]} \,,
\label{average}
\end{equation}
where the action $F[Q]$ is
\begin{eqnarray}
F[Q] &  =  &  \int d{\bf r}\, {\rm str}\bigg\{\left(\frac{\pi
N(\Omega^2)}{2}\right)^2\Delta Q^2-
\label{F}\\
&  &  \frac{1}{2}\ln\left[-i{\cal H}_0- i\Omega\omega^+
\Lambda+\pi\Delta N(\Omega^2)Q({\bf r})\right] \bigg\} \,. \nonumber
\end{eqnarray}
Minimizing $F[Q]$ gives the saddle point equation:
\begin{eqnarray}
Q &=& \frac{1}{\pi N(\Omega^2)}\, \left\{-i{\cal H}_0-
i\Omega\omega^+ \Lambda+\pi\Delta N(\Omega^2)Q\right\}^{-1}
\nonumber\\
&\equiv& \frac{1}{\pi N(\Omega^2)}\, \mathcal{G}_0 \label{spa}
\end{eqnarray}
In the limit $\Omega\gg
\omega\,, \pi\Delta N(\Omega^2)/\Omega$\,, Eq.~(\ref{spa}) gives the
saddle point as $Q({\bf r})=\Lambda$\,.

So far the derivation is exact. Fluctuations analysis around the
saddle point may be performed for Eqs.~(\ref{average}) and
(\ref{F}). Yet, we could not proceed further and only give the
results here, instead refer the reader to Ref.~\onlinecite{Efetov97}
for all the details. First, after standard procedure the mean field
approximation namely $Q({\bf r})=\Lambda$ gives the averaged
retarded/advanced Green function as
\begin{equation}
\overline {G^{R,A}_{\Omega^2}({\bf r},{\bf
 r}')}=\langle {\bf r} |\{
 \Omega^2+\nabla^2\pm i\pi \Delta N(\Omega^2)\}^{-1}
  |{\bf r}' \rangle \,.
\label{Greenfunctionaverage}
\end{equation}
The imaginary part of the self-energy gives the elastic mean free
path which is
\begin{equation}
l=\frac{\Omega}{\pi\Delta N(\Omega^2)}\,,
 \label{mfp}
\end{equation}
and has the Rayleigh form, i.e., $l\sim \Omega^{-(d+1)}$\,.

Then, with $\Omega l\gg 1$ taken into account the action is
simplified to be $F[Q]=\int d{\bf r}\, {\cal L}[Q]$\,, where (From
now on we set $\nu\equiv \nu(\Omega)$ to shorten formula.)
\begin{equation}
{\cal L}[Q]=\frac{\pi\nu}{8} \, {\rm str}\, [D_0(\partial Q)^2
+2i\omega^+\Lambda Q] \label{action}
\end{equation}
with the bare diffusion constant $D_0=l/d$\,. Here $Q({\bf r}) =
T({\bf r}) \Lambda T^{-1}({\bf r})$ describes Goldstone modes with
$T({\bf r})$ a matrix field taking the value in the coset space
$U(2,2/4)/U(2/2)\times U(2/2)$ reflecting the orthogonal symmetry.
An explicit parametrization of $T$ will be given in the next
section.

\subsection{$Q$-field constraint at the vacuum-medium interface}
\label{coupling}

The action $F[Q]$ obtained above is invariant under the
translational symmetry, which is broken in the presence of the
vacuum-medium interface. The broken translational symmetry may
profoundly affect light propagation. Experience in mesoscopic
physics shows that to take into account the vacuum-medium interface
effect one may impose some appropriate boundary condition on the
$Q$-field in the field-theoretic formalism. However this is a
nontrivial task and in meseoscopic physics investigations so far
have been restricted on interface structures of
quasi-one-dimensional disordered wires and small quantum dots
\cite{Efetov97,Iida90,Zirnbauer95,Zirnbauer94}. In this part we
switch to optical systems and study the vacuum-medium coupling where
the interface may be infinite and bear arbitrary geometry.

\subsubsection{The vacuum-medium coupling action}
\label{couplingaction}

Though the derivation below may be generalized to arbitrary
dimension to simplify discussions we will focus on the
two-dimensional case. Let us suppose an arbitrary curve $C$ which
divides the space $\mathbb{R}^2$ into two disconnected subspaces
${\cal V}_-$ and ${\cal V}_+$\,, i.e., $\mathbb{R}^2 = {\cal
V}_-\cup {\cal V}_+\cup C$ and ${\cal V}_-\cap {\cal
V}_+=\emptyset$\,. We are interested in light propagation in some
subspace say ${\cal V}_+$ described by an effective Green function
${\cal G}^{R,A}_{\Omega^2}({\bf r}\,, {\bf r}')$\,, which is
identical to $G^{R,A}_{\Omega^2}({\bf r}\,, {\bf r}')$ for ${\bf
r}\,, {\bf r}'\in {\cal V}_+$\,. To study such Green functions for
${\bf r}\,, {\bf r}'\in {\cal V}_-$ we introduce auxiliary Green
functions $g^{R,A}_{\Omega^2}({\bf r}\,, {\bf r}')$ satisfying
\begin{eqnarray}
\{
 \Omega_\pm^2- {\hat H} \}\, g^{R,A}_{\Omega^2}({\bf r},{\bf
 r}')&=&\delta({\bf r} -{\bf r}') \,,
\label{auxiliaryGreenfunction}\\
g^{R,A}_{\Omega^2}({\bf r},{\bf
 r}')|_{{\bf r} \, {\rm or}\, {\bf
 r}'\in C}&=&0\,. \nonumber
\end{eqnarray}
Then our starting point is
the following theorem due to Zirnbauer \cite{Zirnbauer95} and refined by Efetov \cite{Efetov97}, which
was originally established for description of coupling between leads and
mesoscopic devices. The theorem is stated as follows: (For
the self-contained purpose the proof tailored to the present context is
given in Appendix~\ref{DerivationEffGreen}.) \\

{\it For
${\bf r}\,, {\bf r}'\in {\cal V}_+$ the Green function ${\cal G}^{R,A}_{\Omega^2}({\bf r}\,, {\bf r}')$
solves
\begin{eqnarray}
\left\{
 \Omega_\pm^2- {\hat H} \pm i {\hat B} \right\}\, {\cal G}^{R,A}_{\Omega^2}({\bf r},{\bf
 r}') &=& \delta ({\bf r}-{\bf r}')\,, \label{theorem}\\
\partial_{{\bf n}({\bf r})}
{\cal G}^{R,A}_{\Omega^2}({\bf r},{\bf r}')\big|_{{\bf r}\in C} &=& 0 \,, \quad {\bf r}\in C\,,
\nonumber
\end{eqnarray}
where the normal unit vector ${\bf n}({\bf r})$ at ${\bf r}$ points
to ${\cal V}_+$\,. Here}
\begin{eqnarray}
({\hat B} f) ({\bf r}) &\equiv& \int_C d{\bf r}'\, {\rm Im}\, [B({\bf r},{\bf r}')] f({\bf r}')\,, \nonumber\\
B({\bf r},{\bf r}') &=&
\partial_{{\bf n}({\bf r})}\partial_{{\bf n}({\bf r}')} g^R_{\Omega^2}({\bf r},{\bf
r}') \,, for\, {\bf r}\,, {\bf r}'\in C\,.
\label{effGF6}
\end{eqnarray}
\noindent The effective Hamiltonian for the retarded (advanced)
Green function is ${\hat H} \mp i {\hat B}$\,. Remarkably, it is
non-hermitian due to the escape from ${\cal V}_+$ into ${\cal V}_-$
through $C$\,.

In the present case the vacuum-medium interface namely the curve $C$
is a straight line. To proceed we choose the coordinate system
$(r_\perp\,, z)$ with the $z$ ($r_\perp$)-direction perpendicular
(parallel) to the vacuum-medium interface. The vacuum fills the
space $z<0$ where no dielectric scatterers are available. For
technical reasons we assume that the dielectric scatterers located
at $(r_\perp^i\,, 0)$ ($r_\perp^1< r_\perp^2<\cdots$) are uniformly
(in the statistical sense) distributed with the distance between
nearest scatterers $l_i=r_\perp^i-r_\perp^{i-1}$ order of $l$\,. $C$
is located at $z=0^-$ and ${\cal V}_+$ (${\cal V}_-$) is set to be
the medium (vacuum).

Taking into account the boundary condition specified in
Eq.~(\ref{auxiliaryGreenfunction}) we find that the Green function
$g_{\Omega^2}^R$ is
\begin{eqnarray}
&& g_{\Omega^2}^R(r_\perp,z,r_\perp',z') \label{GFinterface}\\
&=& \frac{1}{\pi^2}\int dk_\perp \int_0^\infty dk
\frac{e^{ik_\perp\,(r_\perp-r_\perp')}\,\sin (kz)\sin
(kz')}{\Omega^2-k_\perp^2-k^2+i0^+} \,. \nonumber
\end{eqnarray}
Upon the substitution of Eq.~(\ref{GFinterface}) into
Eq.~(\ref{effGF6}) the operator ${\hat B}$ is simplified to be
\begin{eqnarray}
&  &  ({\hat B} f)\, (r_\perp) \label{Bdefinition} \\
&  =  &
\int_{|k_\perp|\leq \Omega}\frac{dk_\perp}{2\pi}\int dr_\perp '\,
\sqrt{\Omega^2-k_\perp^2}\, e^{ik_\perp\,(r_\perp-r_\perp')}\,
f(r_\perp ') \,. \nonumber
\end{eqnarray}

Repeating the derivation of Sec.~\ref{NSM} with the effective
Hamiltonian ${\hat H} \mp i {\hat B}$ we arrive again at Eq.~(\ref{F}) except that the action
is modified according to $F[Q] \rightarrow  F[Q] + F_{\rm inter}[Q]$
with
\begin{eqnarray}
F_{\rm inter}[Q]
= -\frac{1}{2}\, {\rm str}\ln\, \left\{1-{\hat B} \Lambda {\cal G}_0\right\} \,. \label{interF}
\end{eqnarray}
Expanding the logarithm and substituting Eq.~(\ref{Bdefinition}) into it
we obtain
\begin{eqnarray}
F_{\rm inter}[Q] =  -\frac{1}{2} \, {\rm str}
\left\{\sum_{n=1}^\infty\, \frac{(-1)^{n+1}}{n} \left[{\hat B}
\Lambda {\cal G}_0\right]^n\right\} \,,
\label{interF1}
\end{eqnarray}
where the supertrace ${\rm str}$ includes the integration over
$r_\perp$\,, and ${\cal G}_0({\bf r},{\bf r}';Q)$ exponentially
decays for $|{\bf r}-{\bf r}'|\gtrsim l$ according to
Eq.~(\ref{spa}). To calculate Eq.~(\ref{interF1}) we introduce, for
arbitrary $i$\,, the auxiliary variable $x_i\equiv
r^i_\perp-r_\perp,\, r_\perp\in [r_\perp^{i-1},\, r_\perp^i]$\,.
Following Ref.~\onlinecite{Efetov97} in the layer $0\leq z\leq l$
the Green function ${\cal G}_0({\bf r},{\bf r}';Q)\equiv {\cal
G}_0(r_\perp^i-x_i, z, r_\perp^{i'}-x_{i'}', z',;Q)$ may be
approximated by
\begin{eqnarray}
{\cal G}_0(r_\perp^i-x_i, z, r_\perp^{i'}-x_{i'}', z',;Q)
=\frac{2\delta_{ii'}}{\pi} \sum_{N\geq 1} \label{G0}\\
\int_0^\infty dk \frac{\varphi^i_{\pi N/l_i}(x_i)\varphi^i_{\pi
N/l_i}(x_i')\cos (kz)\cos
(kz')}{\Omega^2(1+\epsilon^i)-k_\perp^2-k^2+i\pi \Delta N(\Omega^2)
Q_i} \,, \nonumber
\end{eqnarray}
where the longitudinal wave function is determined by the boundary
condition of Eq.~(\ref{theorem}). Here $Q^i$ and $1+\epsilon^i$
stand for the $Q$- and the dielectric field, respectively in the
regime: $[r_\perp^{i-1}\,,r_\perp^i]\times [0\,, l]$\,. They are
considered to be a constant (matrix) since both $Q$ and $\epsilon$
varies over the scale $l$\,. Moreover, the transverse component
$\varphi^i_{k_\perp}$ is defined as
\begin{equation}
\varphi^i_{\pi N/l_i}(x_i)=\sqrt{\frac{2}{l_i}}\, \sin \frac{\pi N
x_i}{l_i} \,.
\label{phi}
\end{equation}

Substituting Eq.~(\ref{G0}) into Eq.~(\ref{interF1}), with the help
of the following identity:
\begin{eqnarray}
\int_0^{l_i}\!\!\!\!\int_0^{l_{i'}}\frac{dx_i dx_{i'}'}{2\pi}\,
e^{ik_\perp\, (-x_i+x_{i'}'+r_\perp^i-r_\perp^{i'})}\,
\varphi^i_{\frac{\pi
N}{l_i}}(x_i)\varphi^{i'}_{\frac{\pi N'}{l_{i'}}}(x_{i'}') \nonumber\\
\approx \frac{\delta_{ii'}\delta_{NN'}}{2}\,
\left[\delta\left(k_\perp-\frac{\pi
N}{l_i}\right)+\delta\left(k_\perp+\frac{\pi
N}{l_i}\right)\right]\qquad
\label{orthogonalrelation}
\end{eqnarray}
we obtain:
\begin{eqnarray}
F_{\rm inter}[Q] = -\frac{1}{2} \, \sum_i \sum_{0<
\frac{k_\perp}{\Omega} \leq 1} {\rm str} \ln \left\{1+
\alpha^i_{k_\perp}\, \Lambda Q_i\right\} \label{interF2}
\end{eqnarray}
with
\begin{eqnarray}
\alpha^i_{k_\perp}=\sqrt\frac{\Omega^2-k_\perp^2}{\Omega^2(1+\epsilon^i)-k_\perp^2}
\,.
\label{alpha}
\end{eqnarray}
Taking the advantage of large channel number $\sum_{0<
\frac{k_\perp}{\Omega} \leq 1}\approx \Omega l_i/\pi$ we may
simplify it to be (The details are given in
Appendix~\ref{simplification}.)
\begin{eqnarray}
F_{\rm inter}[Q]  =  -\frac{1 }{4} \, \sum_i \frac{\Omega l_i }{\pi}\, T_0(i) \, {\rm
str}
\, (\Lambda Q_i)
\label{interFresult}
\end{eqnarray}
by assuming that $\alpha_{k_\perp}^i$ does not depend on $k_\perp$\,, i.e.,
$\alpha_{k_\perp}^i\equiv \alpha^i$\,. Here
\begin{eqnarray}
T_0(i)=\frac{4\alpha^i}{(1+\alpha^i)^2}\leq 1
\label{T}
\end{eqnarray}
is the well known transmission coefficient of electromagnetic wave
\cite{LL}. Passing to the continuum limit: $\sum_i l_i\rightarrow
\int dr_\perp$ we rewrite the action $F_{\rm inter}[Q]$ as
\begin{eqnarray}
F_{\rm inter}[Q]  =  -\frac{\Omega}{4\pi}\, \int dr_\perp \, T_0(r_\perp) \, {\rm
str}
\, [\Lambda Q(r_\perp,z=0)]\,.
\label{interFresult1}
\end{eqnarray}

In the quasi-one-dimensional geometry the summation over $i$ is
suppressed, and the coupling action $F_{\rm inter}[Q]$ namely
Eq.~(\ref{interFresult}) recovers the one obtained previously
\cite{Iida90,Zirnbauer94}. For $d>2$ although to generalize the
derivation above is straightforward, the coupling action may be
obtained by simple physical arguments below. Notice that the
coefficient of Eq.~(\ref{interFresult}) allows a simple physical
explanation \cite{Iida90}: According to
Eq.~(\ref{Greenfunctionaverage}) the (single) photon Green function
decays over the scale $l$\,. Suppose that the medium is partitioned
into boxes of volume $l^d$\,, then the states (denoted as $\mu$) in
different boxes are uncorrelated. The box states neighboring to the
interface may be translated into the vacuum state (denoted as
$a$)--so-called lead channel in the terminology of mesoscopic
physics. The coupling strength is $\propto \sum_\mu\, W_{a\mu}
W_{\mu a}$ with $W_{a\mu}$ the scattering matrix element, which
scales as $l^{d-1}/A$ with $A$ the interface area. Thus, although
the total channel number is $\propto A\Omega^{d-1}$\,, the number of
channels to which the interface box state is transmitted (denoted as
$N_d$) is much smaller $N_d \sim A\Omega^{d-1} \times l^{d-1}/A =
(\Omega l)^{d-1}$\,. More precisely, $N_d$ may be found to be
\begin{equation}
N_d = \frac{l^{d-1}}{(2\pi)^{d-1}}\, \frac{2\pi^{(d-1)/2}}{\Gamma \left(\frac{d-1}{2}\right)}\int_0^\Omega dk_\perp\, k_\perp^{d-2}\,.
\label{channelnumber}
\end{equation}
For $d=2$ this gives $N_2=\Omega l/\pi$ namely the coefficient of
Eq.~(\ref{interFresult})\,. For arbitrary $d$ with the replacement
of $N_2\rightarrow N_d$ (and $l\rightarrow l_i$) in
Eq.~(\ref{interFresult}) the vacuum-medium action becomes
\begin{eqnarray}
{\tilde F}_{\rm inter}[Q]  &=& -\frac{{\tilde N}_d }{4} \, \sum_i \frac{(\Omega l_i)^{d-1} }{\pi}\, T_0(i) \, {\rm
str}
\, (\Lambda Q_i) \nonumber\\
&=& -\frac{{\tilde N}_d \Omega^{d-1}}{4} \, \int d{\bf r}\delta_C\,
T_0({\bf r})\, {\rm str} [Q({\bf r})\Lambda]
\label{interFresult2}
\end{eqnarray}
with
\begin{equation}
{\tilde N}_d = \frac{1}{(2\pi)^{d-1}}\, \frac{\pi^{(d-1)/2}}{\frac{d-1}{2}\Gamma \left(\frac{d-1}{2}\right)}\,.
\label{channelnumber1}
\end{equation}
In the last equality of Eq.~(\ref{interFresult2}) we again pass to
the continuum limit, and the operator $\delta_C$ is defined as $\int
d{\bf r}\delta_C \, f({\bf r})\equiv \int d{\bf r}_\perp f({\bf
r}_\perp,z=0)$\,.

\subsubsection{Boundary condition}
\label{boundarycondition}

We then come to derive the boundary condition satisfied by $Q$\,.
For this purpose we employ the so-called boundary Ward identity
\cite{Altland98}. It states that an arbitrary local observable say
$P({\bf r})$ (with ${\bf r}$ inside the medium), which is expressed
in terms of the average of the functional ${\cal P}[Q({\bf r})]$
namely
\begin{equation}
P({\bf r})\equiv\int D[Q] {\cal P}[Q({\bf r})]\, e^{-\int_{z\in
\mathbb{R}^+}d{\bf r}{\cal L}[Q({\bf r})]-{\tilde F}_{\rm inter}[Q]}\,,
\label{P}
\end{equation}
must be invariant under an
infinitesimal boundary
rotation below:
\begin{eqnarray}
Q  &  \rightarrow  &  e^{-R}Qe^R\approx Q-[R,Q] \,,\label{rotation}\\
R
&  =  &
\left(
  \begin{array}{cc}
    0 & {\cal R}({\bf r}_\perp,z=0) \\
    {\bar {\cal R}}({\bf r}_\perp,z=0) & 0 \\
  \end{array}
\right)^{ar}\otimes \mathbf{1}^{bf} \otimes \mathbf{1}^{tr}\,.
\nonumber
\end{eqnarray}

Notice that the boundary rotation alters neither ${\cal P}[Q({\bf
r})]$ nor ${\cal L}[Q({\bf r})]$ for ${\bf r}$ inside the medium.
The boundary Ward identity then demands $\delta {\tilde F}_{\rm inter}\equiv
0$\,, i.e.,
\begin{eqnarray}
&& \delta {\tilde F}_{\rm inter} \label{actionchange}\\
&=& \int d{\bf r}\delta_C {\rm str}\left\{R\left(\frac{\pi \nu
D_0}{2} Q\partial_z Q + \frac{{\tilde N}_d \Omega^{d-1}}{4} \,T_0
[Q,\Lambda]\right)\right\} \nonumber\\
&=& 0 \,. \nonumber
\end{eqnarray}
As ${\cal R}\,, {\bar {\cal R}}$ are arbitrary this requires
\begin{equation}
\left({\tilde l}\, Q\partial_z Q + T_0
[Q,\Lambda]\right)_\perp\Big|_{z=0}=0 \label{BC}
\end{equation}
to be met, where the subscript $\perp$ stands for the offdiagonal
component in the retarded/advanced sector (thereby anticommunting
with $\Lambda$) and ${\tilde l}=2\pi \nu D_0/({\tilde N}_d
\Omega^{d-1})$\,.

Eq.~(\ref{BC}) is the field-theoretic version of the radiative
boundary condition \cite{Lagendijk00,Agranovich91,Lagendijk89}. As
we will show in Sec.~\ref{WLsemiinfinite}, it describes that the
low-energy coherent dynamics penetrates into the vacuum with the
extrapolation length
\begin{equation}
\zeta=\frac{\tilde l}{2T_0}=\frac{\pi \nu D_0}{{\tilde N}_d \Omega^{d-1}}\, \frac{1}{T_0}\,.
 \label{extrapolationlength}
\end{equation}
Notice that it is proportional to the inverse transmission
coefficient in agreement with Ref.~\onlinecite{Agranovich91}. For
$d=3$ in the case of perfect transmission, i.e., $T_0=1$ the
extrapolation length is $\zeta=\frac{2}{3}\, l$ in agreement with
Ref.~\onlinecite{Lagendijk00}, and is closed to the one that
obtained by solving Milne equation \cite{Davison} which gives
$\zeta=0.7 l$\,. Traditionally the radiative boundary condition is
imposed to diffusion equation to mimics the leakage at the interface
\cite{Lagendijk00,Agranovich91,Lagendijk89,Oppenheim72} and is
justified for one-dimensional discrete random walk
\cite{Oppenheim72}.

\section{Two-dimensional renormalization effects of infinite medium bar}
\label{renormal}

In the rest of this paper we will apply the supersymmetric
field-theoretic formalism to the semi-infinite two-dimensional
medium bar (with the width $a\gg l$), where in the transverse
($\rho$-) direction the photon motion is confined. The purpose of
this section is two-fold: On the physical side, we wish to explore
how a finite width affects localization in the bulk which differs
inessentially from localization in an infinite bar. Accordingly,
through out this section the action reads out as
$F[Q]=\int_{-\infty}^\infty dz\int_0^a d\rho\, {\cal L}[Q]$\,. On
the technical side, by presenting some details we wish to address
the difference of calculations between semi-infinite and infinite
bar, which originates at the fact that in the former system the
translational symmetry of low-energy modes, i.e., the $Q$-field is
broken.

Following the standard strategy we factorize the $T$-field into the
slow and fast mode in terms of $T=T_>T_<$\,, where rotations $T_>$
($T_<$) involve spatial fluctuations on short (large) scales.
Substituting it into the action we then obtain:
\begin{eqnarray}
F[Q] &  =  &  \frac{\pi\nu}{8}\int_{-\infty}^\infty \!\!\!\!dz
\!\!\int_0^a \!\!\!\! d\rho\, {\rm str} \, \big\{D_0\big((\partial
Q_>)^2 + \label{action1} \\
&  &   4Q_>\partial Q_>\cdot \Phi + [\Phi, Q_>]^2\big) + 2i\omega^+
Q_> T_<^{-1}\Lambda T_< \big\}\,, \nonumber
\end{eqnarray}
where $Q_>=T_>\Lambda T_>^{-1}$ and $\Phi=T_<^{-1}\partial T_<$\,.
Integrating out $Q_>$ results in an effective action of $Q_<$\,.

\subsection{Parametrization of fast modes}
\label{parametri}

To work out the strategy outlined above we set $T_>= 1+iW_>$ with
$W_>$ parameterized by
\begin{eqnarray}
W_>=\left(
\begin{array}{cc}
  0 & B_> \\
  {\bar B_>} & 0 \\
\end{array}
\right)^{ar} \,.
\label{parametrization}
\end{eqnarray}
Since photons are confined the current vanishes at $\rho=0\,, a$\,,
i.e., $\partial_\rho W_>({\bf r})|_{\rho=0\,, a}=0$\,. We may thus
introduce the Fourier transformation
\begin{eqnarray}
W_>({\bf r}) = \frac{2}{a}\int_{|k|\geq k_0}
\frac{dk}{2\pi}\sum_{n\geq n_0}
 W_{k,n\pi/a} \, e^{ikz}\, \cos\frac{n\pi \rho}{a} \,,\nonumber\\
 \label{WFourier}
\end{eqnarray}
where $k_0$ and $\pi n_0/a$ are ultraviolet cut-off of longitudinal
and transverse wave number, respectively. In
Eq.~(\ref{parametrization}) the matrix $B_>$ has the structure as
\begin{eqnarray}
B_> = \left(
\begin{array}{cc}
  a & i\sigma \\
 \eta & ib \\
\end{array}
\right)^{bf} \label{B}
\end{eqnarray}
with
\begin{eqnarray}
a = \left(
\begin{array}{cc}
  a_1 & a_2 \\
 -a_2^* & a_1^* \\
\end{array}
\right)^{tr}\,, \qquad b = \left(
\begin{array}{cc}
  b_1 & b_2 \\
 b_2^* & b_1^* \\
\end{array}
\right)^{tr}\,, \nonumber\\
\sigma = \left(
\begin{array}{cc}
  \sigma_1 & \sigma_2 \\
 -\sigma_2^* & -\sigma_1^* \\
\end{array}
\right)^{tr}\,, \qquad  \eta = \left(
\begin{array}{cc}
  \eta_1 & \eta_2 \\
 \eta_2^* & \eta_1^* \\
\end{array}
\right)^{tr}\,,
\label{absigmaeta}
\end{eqnarray}
where $a$'s, $b$'s ($\sigma$'s, $\eta$'s) are complex bosonic
(Grassmann) numbers, and the charge conjugation transformation of a
matrix $M$ is defined as
\begin{eqnarray}
{\bar M}=C_0 M^{\rm T} C_0^{\rm T} , \quad C_0 = \left(
\begin{array}{cc}
  -i\tau_2 & 0 \\
 0 & \tau_1 \\
\end{array}
\right)^{bf} \,.
\label{C}
\end{eqnarray}
Straightforward calculations justify the useful identity:
$\overline{M_1 M_2} = \bar{M_2}\bar{M_1}$\,.

Importantly, $W_>$ satisfies the following relation:
\begin{eqnarray}
W_> = KW_>^\dagger K\,,\quad
 K = \left(
                            \begin{array}{cc}
                              1 & 0 \\
                              0 & k \\
                            \end{array}
                          \right)^{ar}\,,
                          \label{Wrelation}
\end{eqnarray}
which, as shown in Appendix~\ref{Wsymmetry}, enforces the invariance
of $W_>$ under the charge conjugation, i.e., ${\bar W}_>=W_>$\,.

\subsection{One-loop renormalization}
\label{renormalaction}

Now we study the one-loop renormalization. In doing so we expand
$Q_>$ to quadratic order in $W_>$\,. Consequently the action
separates into three contributions:
\begin{equation}
F=F_S+F_F+F_{SF} \,,
\label{actionSF}
\end{equation}
where the slow mode action is
\begin{eqnarray}
F_S= \frac{\pi\nu}{8}\int_{-\infty}^{\infty} \!\!\!\!dz \!\!\int_0^a
\!\!\!\! d\rho\, {\rm str} [D_0\, (\partial Q_< )^2
+2i\omega^+\Lambda Q_<] \label{Saction}
\end{eqnarray}
with $Q_<=T_<\Lambda T_<^{-1}$\,, the fast mode action is
\begin{eqnarray}
F_F= \frac{\pi\nu D_0}{2} \int_{-\infty}^{\infty} \!\!\!\!dz
\!\!\int_0^a \!\!\!\! d\rho\, {\rm str} \, (\partial W_> )^2 \,,
\label{Faction}
\end{eqnarray}
and the slow-fast mode coupling is described by the action:
\begin{widetext}
\begin{eqnarray}
 F_{SF}&  =  &  \pi\nu \int_{-\infty}^{\infty} \!\!\!\!dz
\!\!\int_0^a \!\!\!\! d\rho\,  {\rm str} \bigg \{
D_0\left([W_>,\partial W_>]\cdot \Phi -(\Phi\Lambda W_>)^2 - (\Phi
\Lambda) ^2 W_>^2 -i \partial W_>\cdot \Phi - i(\Phi\Lambda)^2 W_>
\right) - \nonumber\\
&  &  \frac{i\omega^+}{2} \left(i W_>\Lambda T_<^{-1}\Lambda T_< +
W_>^2 \Lambda T_<^{-1}\Lambda T_< \right)\bigg\} \,.
\label{SFaction}
\end{eqnarray}

In order to calculate the general average with respect to the fast
mode action: $\langle\cdots\rangle_{F}\equiv \int D[W_>]\,
(\cdots)e^{-F_F}$ we employ the contraction rule below:
\begin{eqnarray}
2\pi\nu \, \langle W_>({\bf r}) {\bar M} W_>({\bf r}')\rangle_F &  =
& -{\cal D}_F({\bf r}\,, {\bf r}')\, M \,, \qquad M=\left(
                 \begin{array}{cc}
                   0 & M^{12} \\
                   M^{21} & 0 \\
                 \end{array}
\right)^{ar}  \,, \nonumber\\
4\pi\nu \, \langle W_>({\bf r}) N W_>({\bf r}')\rangle_F &  =  &
{\cal D}_F({\bf r}\,, {\bf r}')\, \left({\rm str} N - \Lambda {\rm
str} \Lambda N\right)\,, \qquad N=\left(
                 \begin{array}{cc}
                   N^{11} & 0 \\
                   0 & N^{22} \\
                 \end{array}
\right)^{ar}  \,, \nonumber\\
2\pi\nu \, \langle {\rm str} [W_>({\bf r}) M_1]\, {\rm str}
[W_>({\bf r}') M_2] \rangle_F & = & {\cal D}_F({\bf r}\,, {\bf
r}')\, {\rm str}[(M_1+{\bar M}_1) M_2] \,, \qquad M_{1,2}=\left(
                 \begin{array}{cc}
                   0 & M^{12}_{1,2} \\
                   M^{21}_{1,2} & 0 \\
                 \end{array}
\right)^{ar} \label{Wcontraction}
\end{eqnarray}
\end{widetext}
with the fast mode propagator
\begin{eqnarray}
  &  &  {\cal D}_F({\bf r}\,, {\bf r}') \label{fastpropagator} \\
&  =  & \frac{2}{a}\int_{|k|\geq k_0} \frac{dk}{2\pi}\, \sum_{n\geq
n_0}\, \frac{\cos\frac{n\pi(\rho-\rho')}{a}\, e^{i
k(z-z')}}{D_0\left[\left(\frac{n\pi}{a}\right)^2+k^2\right]}
\nonumber
\end{eqnarray}
and the Wick theorem. Notice that we assume that the fast mode
propagator does not depend on the transverse center-of-mass
coordinate, i.e., $(\rho+\rho')/2$ and the self-average over this
variable has been performed.

Performing the average of $F_{SF}$ we obtain an effective action
$F_{\rm eff}[Q_<]= F_S+ \langle F_{SF} \rangle_F$\,, where
\begin{equation}
\langle F_{SF}\rangle_F=-\frac{I_0}{8}\int_{-\infty}^{\infty}
\!\!\!\!dz \!\!\int_0^a \!\!\!\! d\rho\, {\rm str}\, (\partial
Q_<)^2 \label{SFactionresult}
\end{equation}
with
\begin{equation}
I_0=\frac{2}{a}\, \int_{|k|\geq k_0} \,\frac{dk}{2\pi} \sum_{n\geq
n_0} \, \frac{1}{\left(\frac{n\pi}{a}\right)^2+k^2} \,. \label{I0}
\end{equation}
Eqs.~(\ref{SFactionresult}) and (\ref{I0}) show that the one-loop
renormalization results in the weak localization correction to the
bare diffusion constant $D_0$\,:
\begin{eqnarray}
F_{\rm eff}[Q_<] &=& \frac{\pi\nu}{8}\, \int_{-\infty}^{\infty}
\!\!\!\!dz \!\!\int_0^a \!\!\!\! d\rho\, {\rm str} \{[D_0+\delta
D^{(2)}]\, (\partial Q_< )^2 +\nonumber\\
&& 2i\omega^+\Lambda Q_<\} \label{effaction}
\end{eqnarray}
with the two-dimensional weak localization correction $\delta
D^{(2)}=-I_0/(\pi\nu D_0)$\,.

\subsection{Dimensional crossover of effective action}
\label{crossover}

In the high-frequency region, i.e., $\omega \gg D_0/a^2$\,, the
condition: $k_0 a/\pi\,, n_0 \sim \sqrt{\omega a^2/D_0}\gg 1$ is
met. The photon motion is thus two-dimensional described by the
action, Eq.~(\ref{effaction}). Furthermore, since $\Omega l\gg 1$
the two-dimensional weak localization correction $\delta D^{(2)}$ is
much smaller than $D_0$\,, the photon motion thereby is diffusive.

In the low-frequency region, i.e., $\omega \lesssim D_0/a^2$ one may
further enforce $k_0=0$ and $n_0=1$ and thereby obtain
a quasi-$1$D effective action of $\breve{Q}\equiv Q_<(z)$ which is
homogeneous in the transverse direction:
\begin{eqnarray}
F_{\rm eff}[\breve{Q}] &=& \int_{-\infty}^{\infty} dz {\cal L}_{\rm
eff}[\breve{Q}]\,, \nonumber\\
{\cal L}_{\rm eff}[\breve{Q}] &=& \frac{\pi\nu a}{8}\,
 {\rm str}\, [D_{\rm eff}\, (\partial_z \breve{Q})^2 +2i\omega^+\Lambda \breve{Q}] \,,
\label{action2DWL}
\end{eqnarray}
where the renormalized diffusion constant is
\begin{equation}
D_{\rm eff} = D_0\left(1-\frac{1}{\pi\nu D_0}\sum_{n\geq
1}\frac{1}{n\pi}\right) \,. \label{Deff}
\end{equation}
Note that in the above the second term suffers logarithmic
divergence which, as usual, may be regularized by introducing the
upper cut-off $N$ which is order of $\sim a/(\pi l)$\,. As a result,
\begin{equation}
D_{\rm eff} \approx D_0\left(1-\frac{1}{\pi^2 \nu D_0}\ln
\frac{a}{\pi l}\right) \,. \label{Deffapprox}
\end{equation}

Thus, in the low-frequency region: $\omega\lesssim D_0/a^2$ the
system is quasi-one-dimensional provided that the bar width
satisfies $a\ll l e^{\pi^2\nu D_0}$\,. For wider bar the system
display two-dimensional strong localization which is beyond the
scope of present perturbative analysis.

\section{Weak localization in semi-infinite transparent medium bar}
\label{WLsemiinfinite}

The discussions of Sec.~\ref{renormal} break down in the
semi-infinite geometry due to the absence of the translational
symmetry. In this section and the next we turn to study the
vacuum-medium interface effect on wave interference.

\subsection{Simplified boundary condition}
\label{simplifyBC}

In order to explore the physics implied by the boundary constraint
Eq.~(\ref{BC}) let us parameterize $T$ in the same way as
Eqs.~(\ref{parametrization}), (\ref{B}) and (\ref{absigmaeta}). (To
distinct notations from those of Sec.~\ref{renormal} we eliminate
all the subscript $>$\,.) With the substitution of the
parametrization and keeping Eq.~(\ref{BC}) up to the first order in
$W$ we obtain:
\begin{eqnarray}
\left(
                            \begin{array}{cc}
                              0 & ({\tilde l}\partial_z-2T_0)B \\
                              ({\tilde l}\partial_z-2T_0){\bar B} & 0 \\
                            \end{array}
                          \right)^{ar}=0
\label{BCsimplify1}
\end{eqnarray}
implying $B\,, {\bar B}\sim e^{z/\zeta}\,, z<0$ with $\zeta={\tilde
l}/(2T_0)$\,. Hence the low-energy Goldenstone modes penetrate into
the vacuum of a depth $\zeta$ then exponentially decays. That is,
the optical paths underlying coherent multiple scattering do not
cross the line located at $z=-\zeta$\,.

From now on we assume that the interface is almost
transparent namely $T_0$ closed to $1$\,. In this case $\zeta=\frac{2}{3}\, l$\,.
Since the mean free path $l$
is much smaller than any other macroscopic scale we may safely
assume that the crossing line where $W$ vanishes coincides with $C$\,.
Consequently the boundary constraint
Eq.~(\ref{BC}) is simplified as
\begin{equation}
Q|_{z=0}=\Lambda \,.
\label{BCsimplify2}
\end{equation}
It is the action: $ F[Q]=\int_0^{\infty} dz\int_0^a d\rho \, {\cal
L}[Q] $ with the $Q$-field subject to this boundary constraint that
we will use in the rest of this paper.

\subsection{Bare diffusive propagator}
\label{classicaldiffu}

Expanding $Q$ in terms of $W$ gives
\begin{eqnarray}
F[Q]=F_{2}[W] + F_{4}[W] + \cdots\,, \label{actionexpansion}
\end{eqnarray}
where the Gaussian action:
\begin{eqnarray}
F_{2}[W] =  \frac{\pi\nu}{2} \!\!\int_{0}^{\infty}\!\!\!\! dz
\!\!\int_0^a \!\!\!\! d\rho \, {\rm str} \, [D_0 (
\partial W )^2
-i\omega W^2 ]
\label{effectiveaction}
\end{eqnarray}
and
\begin{eqnarray}
F_{4}[W] &=& \frac{\pi\nu}{2} \int_0^{\infty}\!\!\!\! dz
\!\!\int_0^a \!\!\!\! d\rho\,\big\{ -2 D_0\, {\rm str}\,
\left[(\partial W)^2 W^2\right] \nonumber\\
&&\qquad\qquad\qquad \quad \,\,+ i\omega\, {\rm str}\, W^4 \big\}
\,. \label{F4}
\end{eqnarray}

From Eq.~(\ref{effectiveaction}) immediately we obtain the same
contraction rules as Eq.~(\ref{Wcontraction})
except making the replacement:
\begin{eqnarray}
{\cal D}_F ({\bf r},{\bf r}') \rightarrow {\cal D} ({\bf r},{\bf
r}';\omega) \,, \label{replacement}
\end{eqnarray}
where the propagator solves the diffusion equation:
\begin{eqnarray}
\left(-D_0 \partial^2-i\omega\right) {\cal
D} ({\bf r},{\bf r}';\omega) &  =  &  \delta({\bf r}-{\bf r}')\,,\nonumber\\
{\cal D}|_{{\bf r}\, {\rm or}\, {\bf r}' \in C}  &  =  &  0 \,.
\label{diffusion}
\end{eqnarray}
The boundary condition above is inherent from the constraint
Eq.~(\ref{BCsimplify2}) which imposes $W({\bf r})|_{{\bf r}\in
C}=0$\,.

Keeping the prefactor of Eq.~(\ref{DC}) up to quadratic term we
obtain the leading cooperon propagator:
\begin{eqnarray}
{\cal Y}^{\rm C}_{(0)} ({\bf r},{\bf r}' ;\omega
 )
& = & \left[\frac{\pi N(\Omega^2)}{2}
  \right]^2 \, \big\langle
 {\rm str} \big\{k (1+\Lambda)(1- \tau_3) W({\bf r})
 \nonumber\\
&  &   \qquad\qquad \quad \times k (1-\Lambda)(1+ \tau_3) W({\bf
r}') \big\} \big\rangle_{F_2}
\nonumber\\
&  =  & \frac{2\pi\nu}{\Omega^2} \, {\cal D} ({\bf r},{\bf
r}';\omega)\,.
\label{cooperon}
\end{eqnarray}
It is easy to see that the propagator above preserves the symmetry
${\cal Y}^{\rm C}_{(0)} ({\bf r},{\bf r}' ;\omega)={\cal Y}^{\rm
C}_{(0)} ({\bf r}',{\bf r};\omega)$ inherent from Eq.~(\ref{DC}).
Eq.~(\ref{cooperon}) is traditionally obtained by summing up all the
ladder diagrams and imposing appropriate boundary condition
\cite{Golubentsev}.

\subsection{Weak localization correction}
\label{YConeloop}

We proceed to calculate the one-loop correction to the bare
propagator ${\cal Y}^{\rm C}_{(0)}$\,. For this purpose we keep the
$W$-expansion up to the quartic terms for both the prefactor and the
action which, after straightforward calculations, gives the cooperon
as ${\cal Y}^{\rm C} \approx {\cal Y}^{\rm C}_{(0)} + \delta {\cal
Y}^{\rm C}$ with
\begin{widetext}
\begin{eqnarray}
  \delta {\cal Y}^{\rm C} ({\bf r},{\bf r}'
  ;\omega
  ) &  =  &  -\left[\frac{\pi N(\Omega^2)}{2}
  \right]^2 \, \big\langle {\rm str} \left\{k (1+\Lambda)(1- \tau_3) W^3({\bf r})
k (1-\Lambda)(1+ \tau_3) W ({\bf r}')\right\} \nonumber\\
&  & + \, {\rm str} \left\{k (1+\Lambda)(1- \tau_3) W({\bf r}) k
(1-\Lambda)(1+ \tau_3) W^3 ({\bf r}')\right\}
\nonumber\\
&  & - \, {\rm str} \left\{k (1+\Lambda)(1- \tau_3) W^2({\bf r}) k
(1-\Lambda)(1+ \tau_3) W^2 ({\bf r}')\right\} \nonumber\\
&  & + \, {\rm str} \left\{k (1+\Lambda)(1- \tau_3) W({\bf r}) k
(1-\Lambda)(1+ \tau_3) W ({\bf r}')\right\} F_{4}[W]
\big\rangle_{F_2}\,.
 \label{cooperonSL1}
\end{eqnarray}
First, it is easy to show that the third term in the right hand side
of Eq.~(\ref{cooperonSL1}) vanishes. Second, as shown in
Appendix~\ref{cancelation} the first two terms partly cancel the
last term. Eventually Eq.~(\ref{cooperonSL1}) is reduced into
\begin{eqnarray}
  \delta {\cal Y}^{\rm C} ({\bf r},{\bf r}'
  ;\omega
  ) &  =  & - \left[\frac{\pi N(\Omega^2)}{2}
  \right]^2 \, (\pi\nu D_0) \int_0^{\infty}\!\!\!\! dz_1 \!\!\int_0^a
\!\!\!\! d\rho_1\, \big\langle  {\rm str} \left\{k (1+\Lambda)(1-
\tau_3) W({\bf r}) k (1-\Lambda)(1+ \tau_3) W ({\bf r}')\right\}
\times
\nonumber\\
&  &  {\rm str}\, \big[\partial^2 \overbrace{W \,({\bf r}_1) W({\bf
r}_1)W}\, ({\bf r}_1) W({\bf r}_1)+\left(\partial W ({\bf r}_1)\,
W({\bf r}_1)\right)^2+\left(\partial W({\bf r}_1)\right)^2 W^2({\bf
r}_1)\big] \big\rangle_{F_2} \,,
 \label{cooperonSL2}
\end{eqnarray}
\end{widetext}
where the overbrace fixes the contraction and the derivative acts
only on the nearest $W$\,. We remark that $\delta {\cal Y}^{\rm C}$
vanishes when ${\cal Y}_{(0)}^{\rm C}$ is spatially homogeneous,
which is a reflection of the flux conservation law or Ward identity
at the one-loop level. Notice that $\delta {\cal Y}^{\rm C}$
vanishes if either ${\bf r}$ or ${\bf r}'$ belongs to the interface
$C$\,. Such property is inherent from Eq.~(\ref{DC}) which vanishes
upon sending either $Q({\bf r})$ or $Q({\bf r}')$ to $\Lambda$\,.
Using the contraction rules and integral by parts we further reduce
Eq.~(\ref{cooperonSL2}) into
\begin{eqnarray}
\delta {\cal Y}^{\rm C} ({\bf r},{\bf r}'
  ;\omega
  ) &  =  &  \frac{2 D_0}{\Omega^2} \, \int_0^{\infty}\!\!\!\! dz_1 \!\!\int_0^a
\!\!\!\! d\rho_1\, {\cal D} ({\bf r}_1,{\bf r}_1;\omega) \times \nonumber\\
&  &  \partial_{{\bf r}_1} {\cal D} ({\bf r},{\bf r}_1;\omega) \,
\partial_{{\bf r}_1} {\cal D} ({\bf r}',{\bf r}_1;\omega)
\label{cooperonSL3}
\end{eqnarray}
after tedious but straightforward calculations. Notice that $\delta
{\cal Y}^{\rm C}$ preserves the symmetry: $\delta {\cal Y}^{\rm C}
({\bf r},{\bf r}' ;\omega ) = \delta {\cal Y}^{\rm C} ({\bf r}',{\bf
r} ;\omega )$\,.

\subsection{Local diffusion equation}
\label{localdiffusion1}

Eq.~(\ref{cooperonSL3}) justifies that ${\cal Y}^{\rm C} = {\cal
Y}^{\rm C}_{(0)}+\delta {\cal Y}^{\rm C}$ solves the following local
diffusion equation:
\begin{eqnarray}
\left\{ -\partial D({\bf r};\omega) \,
\partial -i\omega \right\} {\cal Y}^{\rm C} ({\bf r},{\bf
r}';\omega) &  =  &
\delta({\bf r}-{\bf r}')\,, \nonumber\\
{\cal Y}^{\rm C}|_{{\bf r}\in C} &  =  &  0 \label{localdiffusion}
\end{eqnarray}
at the one-loop level. Here $D({\bf r};\omega) = D_0 + \delta D({\bf
r};\omega)$ with the weak localization correction
\begin{equation}
\delta D({\bf r};\omega) = -\frac{D_0}{\pi\nu}\, {\cal D}({\bf
r},{\bf r};\omega)\,. \label{WL}
\end{equation}
The local diffusion equation differs from the traditional one in
that the diffusion coefficient is position-dependent. It may be
amounted to incompletely developed constructive interference between
two counter-propagating optical paths--which leads to the weak
localization--near the boundary. Indeed, although deep inside the
medium $\delta D({\bf r};\omega)$ saturates recovering the bulk weak
localization, at the interface it vanishes, i.e.,
\begin{equation}
\delta D({\bf r};\omega)|_{z=0} =0\,.
\label{WL1}
\end{equation}
Importantly, this is contrary to the theoretical proposal of
Ref.~\onlinecite{Berkovits87} which claims that wave interference
democratically renormalizes the diffusion constant appearing in both
the diffusion equation in the bulk and the radiative boundary
condition at the interface.

Here several remarks are in order: (i) Higher order loop corrections
preserve Eq.~(\ref{localdiffusion}). They affect the local diffusion
equation by introducing higher order weak localization corrections
which are also position-dependent. This peculiar property reflects
the photon number conservation law and is protected by Ward
identity. (ii) In the presence of internal reflection namely
$T_0({\bf r})$ (far) below $1$\,, (i) is no longer applicable
because the simplification namely Eq.~(\ref{BCsimplify2}) breaks
down due to large extrapolation length. In fact,
Ref.~\onlinecite{Lagendijk00} falls into this case. (iii) The
concept of local diffusion originally introduced in
Ref.~\onlinecite{Lagendijk00} at the static limit, i.e.,
$\omega\rightarrow 0$ together with its dynamic generalization
\cite{Skipetrov04,Skipetrov06} is now justified at the perturbative
level.

\section{Static limit of local diffusion equation}
\label{localdiffueq}

In this section we study the static limit: $\omega\rightarrow 0$
(For this reason below we suppress the argument $\omega$ in all the
formulae.) of the local diffusion equation namely
Eq.~(\ref{localdiffusion}) for a bar with the width satisfying $l\ll
a \ll l e^{\pi^2 \nu D_0}$\,. In particular we will explicitly
calculate the weak localization correction Eq.~(\ref{WL}), and study
its effects on the coherent backscattering phenomenon.

\subsection{Quasi-$1$D massive local diffusion equation}
\label{localdiffueq1D}

In the static limit the weak localization correction Eq.~(\ref{WL})
becomes self-averaged over the center-of-mass $\rho$ and thereby is
$\rho$-independent. That is,
\begin{equation}
\delta D(z) = -\frac{D_0}{\pi\nu a}\, \int_0^a \!\!\!\! d\rho\,
{\cal D}(z,\rho,z,\rho) \,.
\label{WLselfaverage}
\end{equation}

Substituting Eq.~(\ref{WLselfaverage}) into
Eq.~(\ref{localdiffusion}) we find that ${\cal Y}^{\rm C}({\bf
r}',{\bf r})$ depends on $\rho-\rho'$\,, but not on the
center-of-mass $(\rho+\rho')/2$\,. Therefore, we may introduce the
Fourier transform:
\begin{eqnarray}
&  &  {\cal Y}^{\rm C}(z,z',\rho-\rho') \label{cooperonFourier}\\
&  \equiv  &  \frac{1}{2}\, {\cal Y}^{\rm C}_0(z,z')+\sum_{n\geq
1}\, {\cal Y}^{\rm C}_{\frac{\pi n}{a}}(z,z') \cos\frac{\pi n
(\rho-\rho')}{a}
\nonumber
\end{eqnarray}
and insert it into Eq.~(\ref{localdiffusion}) to obtain
($q_\perp\equiv \frac{n\pi}{a}$)
\begin{eqnarray}
&  &  \left\{ -\partial_z D(z) \,
\partial_z + D(z) q_\perp^2 -i0^+ \right\} {\cal Y}^{\rm C}_{q_\perp} (z,z') =
\delta(z-z') \nonumber\\
&  &
\qquad \qquad \qquad\qquad\qquad {\cal Y}^{\rm C}_{q_\perp} (z=0,z') =  0 \,,
\label{localdiffusion1D}
\end{eqnarray}
where $0^+$ is infinitesimal positive constant.
Eq.~(\ref{localdiffusion1D}) may be considered to be a
quasi-one-dimensional local diffusion equation with a mass $D(z)
q_\perp^2$\,.

\subsection{Dimensional crossover of weak localization}
\label{WL1D}

The weak localization correction Eq.~(\ref{WLselfaverage}) then
becomes
\begin{equation}
\delta D(z) = -\frac{D_0}{\pi\nu a}\, \left[{\cal
D}_0(z,z)+2\sum_{n\geq 1}\, {\cal D}_{\frac{n\pi}{a}}(z,z)
\right]\,. \label{WLselfaverage1}
\end{equation}
Here ${\cal D}_{q_\perp}(z,z')$ satisfies
\begin{eqnarray}
\left\{D_0 \left[-\partial_z^2 + q_\perp^2 \right]-i0^+\right\}
{\cal
D}_{q_\perp}(z,z') &  =  &  \delta(z-z')\,,\nonumber\\
{\cal D}_{q_\perp}(z=0\,, z')  &  =  &  0 \,.
\label{bareproagator}
\end{eqnarray}
It is solved by (with the introduction of ${\tilde
q}_\perp=\sqrt{q_\perp^2-i0^+}$)
\begin{eqnarray}
\frac{{\cal D}_{q_\perp}(z,z')}{\pi\nu a} = \bigg\{\begin{array}{c}
\frac{1}{2\xi }\, \frac{e^{{\tilde q}_\perp z'}-e^{-{\tilde q}_\perp
z'}}{{\tilde q}_\perp} \, e^{-{\tilde q}_\perp z} \,,
                                    z>z' \,,\\
\frac{1}{2\xi}\, \frac{e^{{\tilde q}_\perp z}-e^{-{\tilde q}_\perp
z}}{{\tilde q}_\perp} \, e^{-{\tilde q}_\perp z'} \,,
                                    z<z' \,,
                                  \end{array}
 \label{proagator}
\end{eqnarray}
where $\xi=\pi\nu a D_0$\,. Substituting it into
Eq.~(\ref{WLselfaverage1}) gives
\begin{equation}
\frac{\delta D(z)}{D_0} = -\frac{z}{\xi}-\frac{a}{\xi} \sum_{n\geq
1} \, \frac{1-e^{-2n\pi z/a}}{n\pi} \,, \label{WLresult}
\end{equation}
where the first term is the quasi-one-dimensional contribution, and
the second term is the two-dimensional contribution with $n\pi/a$
standing for the transverse hydrodynamic wave number.

As expected at $z=0$ the weak localization correction $\delta D$
vanishes. Away from the interface i.e., $l\lesssim z\ll a$ it may be
approximated by
\begin{eqnarray}
\frac{\delta D(z)}{D_0} = -\frac{z}{\xi} -\frac{a}{\pi\xi}\,
\ln\frac{z}{l} \label{I1approx}
\end{eqnarray}
as shown in Appendix~\ref{Izdiscussion}. This suggests that in this
region (even in the static limit) the two-dimensional low-energy
motion dominates the weak localization.

The two-dimensional contribution saturates at $z\sim a$\,:
\begin{eqnarray}
\frac{\delta D(z)}{D_0} = -\frac{z}{\xi}-\frac{a}{\xi} \sum_{n\geq
1} \, \frac{1}{n\pi} \,,
\label{I1approx1}
\end{eqnarray}
where the second term is none but the bulk weak localization
correction (see Eq.~(\ref{Deff})) renormalizing the bare diffusion
constant $D_0$\,. With this taken into account Eq.~(\ref{I1approx1})
may be rewritten as
\begin{eqnarray}
\frac{\delta D_{1{\rm D}}(z)}{D_{\rm eff}} = -\frac{z}{\xi_{1{\rm
D}}} \,,
\label{I1approx2}
\end{eqnarray}
where $\delta D_{1{\rm D}}(z)$ stands for the quasi-one-dimensional
weak localization correction, and $\xi_{1{\rm D}}=\pi\nu a D_{\rm
eff}$ is the exact localization length \cite{Efetov97,Efetov83}.
Eq.~(\ref{I1approx2}) agrees with the leading $z/\xi_{1{\rm
D}}$-expansion of the local diffusion coefficient given in
Ref.~\onlinecite{Lagendijk00}. It thereby justifies that for
$z\gtrsim a$ the medium bar displays the quasi-one-dimensional
(interface) weak localization. Indeed, at early times $t\lesssim
D_0/a^2$ incident photons explore a region of size $a^2$ neighboring
to the interface, approaching a uniform distribution in the
transverse direction. At later times they diffuse as in a
quasi-one-dimensional medium. Technically, starting from
quasi-one-dimensional $\sigma$ model by performing the one-loop
calculation one finds Eq.~(\ref{I1approx2}). Importantly, from
Eqs.~(\ref{I1approx}) and (\ref{I1approx2}) we find that within the
boundary layer $z\lesssim \xi$ weak {\it rather than strong}
localization occurs even in the static limit.

\subsection{CBS line shape: crossover from $2$D weak to quasi-$1$D strong localization}
\label{CBS}

In this part we turn to investigate effects of local diffusion on
the CBS line shape. We will consider a medium
illuminated by light of frequency $\Omega$ parallel to the bar, and
calculate the angular resolution of the backscattered light
intensity $\alpha(\theta)$ near the inverse incident direction.
Since the bar is wide enough so that $l \ll a \, (\ll l\, e^{\pi^2
\nu D_0})$ a large parametric region: $\lambda/a \leq \theta
\leq\lambda/l$ is opened. Below we pay particular attention to the
line shape at $0\leq \theta \lesssim \lambda/l$\,. (Notice that the
line shape is symmetric with respect to $\theta=0$\,.)

It is well known that the backscattered light intensity may be
decomposed into the background $\alpha_0$ and the coherent part
$\alpha_c(\theta)$ according to
\begin{equation}
\alpha(\theta)=\alpha_0+\alpha_c(\theta)\,.
\label{albedo}
\end{equation}
Here
\begin{eqnarray}
\alpha_0 & = & \int\!\!\!\!\int d{\bf r}d{\bf r}' \,
e^{-\frac{z+z'}{l} } {\cal Y}^{\rm {D}} ({\bf r},{\bf
r}') \label{DCdefinition}\\
\alpha_c(\theta) & = & \int\!\!\!\!\int d{\bf r}d{\bf r}' \,
e^{-\frac{z+z'}{l} }\, \cos[{\bf q}_\perp\cdot ({\bf r}-{\bf r}')]\,
{\cal Y}^{\rm {C}} ({\bf r},{\bf r}') \,, \nonumber
\end{eqnarray}
where $q_\perp=2\Omega\, \sin (\theta/2)\approx 2\pi\theta/\lambda$
because of $\theta\ll 1$\,, and the overall normalization factor is
omitted. First of all, it is easy to show that
$\alpha_0=\alpha_c(0)$ and therefore only the coherent part
$\alpha_c(\theta)$ which determines the line shape will be studied
below.
Inserting the Fourier transform, namely Eq.~(\ref{cooperonFourier})
into $\alpha_c(\theta)$ we arrive at
\begin{eqnarray}
\alpha_c(\theta) & = & \int_0^{\infty}\!\!\!\!\int_0^{\infty} d z d
z' \, e^{-\frac{z+z'}{l} }\, {\cal Y}^{\rm {C}}_{q_\perp} (z,
z') \nonumber\\
&  \approx  &  l^2\, {\cal Y}^{\rm {C}}_{q_\perp} (l, l) \,,
\label{lineshape}
\end{eqnarray}
where the propagator ${\cal Y}^{\rm {C}}_{q_\perp} (z, z')$ solves
Eq.~(\ref{localdiffusion1D}).

\subsubsection{Signatures of $2$D weak localization}
\label{largeangle}

The interfering optical paths penetrate into the medium of a depth
$\sim q_\perp^{-1}$\,. Let us first study the CBS line shape in the
region: $\pi/a \ll q_\perp\lesssim l^{-1}$\,. Because of the
condition: $l\lesssim q_\perp^{-1} \ll a$ the CBS line shape is
mainly responsible for by photons which diffuse around the
interface, i.e., $l\lesssim z \ll a$ and thereby undergo
two-dimensional weak localization. Indeed, the first term of
Eq.~(\ref{WLresult}) is much smaller due to the condition $|q_\perp
a|/\pi\gg 1$\,. Setting the ultraviolet cutoff $N\sim a/(\pi l)$ (as
Eq.~(\ref{Deffapprox})) we may approximate Eq.~(\ref{WLresult}) by
\begin{equation}
\frac{\delta D(z)}{D_0} \approx -\frac{a}{\xi}
\sum_{aq_\perp/\pi}^{a/(\pi l)} \, \frac{1-e^{-2n\pi z/a}}{n\pi}
\approx \frac{a}{\pi\xi}\,\ln |q_\perp l| \,. \label{WLresultapprox}
\end{equation}

\begin{figure}
\begin{center}
\leavevmode \epsfxsize=8cm \epsfbox{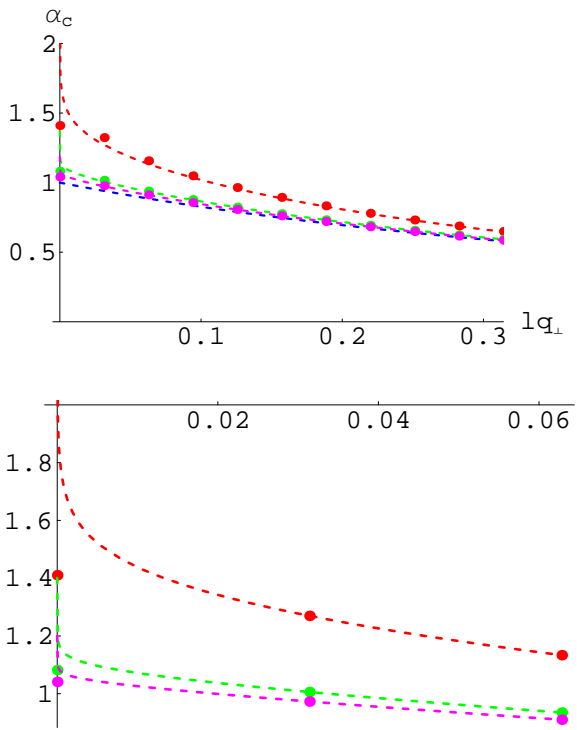}
\end{center}
\caption{(Color online) Coherent backscattering intensity (solid
circle) in unit of $l^3/D_0$ versus $q_\perp=n\pi/a$ for the medium
bar with $a/l=100$\,. Upper panel: The logarithmic
enhancement--dashed line--is cut off at $q_\perp=0$\,. From top to
bottom the parameter $l/\lambda$ is $2$ (red), $10$ (green), $20$
(purple) and $\infty$ (blue). Lower panel: The magnification of the
upper panel for $lq_\perp$ closed to $0$ (The blue curve is not
plotted.). }
  \label{fig}
\end{figure}

With the substitution of such weak localization correction into
Eq.~(\ref{localdiffusion1D}) we find
\begin{eqnarray}
\alpha_c(\theta)   &  =  &   \frac{l^3}{D_0}\, (1-2l q_\perp)\,
\left[1-\frac{a}{\pi\xi}\, \ln |q_\perp l|\right] \,, \nonumber\\
&  & \pi /a \ll q_\perp \lesssim l^{-1} \,.
\label{lineshaperesult1}
\end{eqnarray}
According to Eq.~(\ref{lineshaperesult1}) the conventional
triangular peak described by the factor $1-2l q_\perp$ is enhanced
by a logarithmic factor (dashed line in Fig.~\ref{fig}).
Eq.~(\ref{WLresultapprox}) indicates that in the region: $\pi/a \ll
q_\perp\lesssim l^{-1}$ the local diffusion is of minor importance.
It is the two-dimensional bulk  weak localization that is
responsible for such a logarithmic enhancement. Thus in such region
the scaling theory, still, is applicable.

\subsubsection{Signatures of quasi-$1$D strong localization}
\label{CBSdirection}

At the exact backscattering direction, i.e., $\theta=0$ interfering
optical paths penetrate into the medium bulk: $z\gg \xi$ where
quasi-one-dimensional (bulk) strong localization states are formed.
In contrast to $\pi/a \ll q_\perp\lesssim l^{-1}$ for $q_\perp=0$
the local diffusion plays crucial roles and thus strongly affects
the CBS line shape. Indeed, despite of the nonperturbative nature of
strong localization the backscattering light intensity at
$q_\perp=0$ may be easily found provided that in the region
$z\gg\xi$ the local diffusion equation is still valid
\cite{Lagendijk00}. For $q_\perp=0$ from
Eq.~(\ref{localdiffusion1D}) one may find
\begin{equation}
{\cal Y}_{q_\perp=0}^{\rm C}(l\,, l)=\int_0^l dz \frac{1}{D(z)}
\approx \frac{l}{D_{\rm eff}}\,, \label{returningprobability}
\end{equation}
where in the second equality the substitution of
Eq.~(\ref{I1approx1}) is made. This immediately gives
\begin{equation}
\alpha_c(0)= \frac{l^3}{D_{\rm eff}} \,.
\label{albedoexactdirection}
\end{equation}
Eq.~(\ref{albedoexactdirection}) shows that although due to
two-dimensional bulk weak localization the CBS line shape develops a
logarithmic singularity at $\pi/a \leq q_\perp\lesssim l^{-1}$\,,
the singularity is cut off at $q_\perp=0$ (Fig.~\ref{fig}) where
quasi-one-dimensional strong localization occurs in the bulk.

It is in order to remark that formally there is a region
$q_\perp\xi\lesssim 1$ where the rounding due to the local diffusion
occurs (Yet, the detailed rounding form depends on $D(z)$ in the
region $z\gtrsim \xi$\,, to find which is beyond the present
perturbative treatise.), however, it is unobservable because the
finite bar width renders $\xi^{-1}\ll \pi/a$\,.

Finally we anticipate that the predicted CBS line shape is
qualitatively correct for $l\gtrsim \lambda$\,, though the
analytical result here is obtained for $l\gg \lambda$\,.

\section{Conclusions}
\label{conclusion}

For light propagation in fully disordered media a supersymmetric
field theory is presented. The supersymmetric $\sigma$ model
described by Eq.~(\ref{action}) may be applied to bulk (infinite)
media for studies of optical localization transition. In this
direction it may serve as an alternative to the replica field theory
\cite{John84}. However, the supersymmetric field-theoretic formalism
turns out to be far more powerful as propagation of incident light
in semi-infinite media \cite{Tian07} concerned, which is the subject
of this paper and is closely related to the coherent backscattering
phenomenon.

Differing from infinite medium in the presence of the vacuum-medium
interface $C$ (The interface may bear arbitrary geometry but must be
smooth over the scale of the mean free path.) the supermatrix field
(locally) satisfies the radiative boundary condition Eq.~(\ref{BC}).
Accordingly, the bare diffusion constant acquires a
position-dependent wave interference correction namely
\begin{equation}
D({\bf r}\,; \omega) = D_0+\delta D({\bf r}\,; \omega) \,,
\label{diffusioncoefficient}
\end{equation}
which roots in the incomplete constructive interference (weak
localization) near the interface. Thus, we justify the (static)
local diffusion equation, originally proposed in
Ref.~\onlinecite{Lagendijk00}, as well as its dynamic (i.e.,
$\omega\neq 0$) generalization \cite{Skipetrov04}. Most importantly,
for (almost) transparent interface, i.e., $T_0({\bf r})\approx 1\,,
{\bf r}\in C$\,, the weak localization correction $\delta D({\bf
r}\,; \omega)$ vanishes at the interface. This immediately shows
that the radiative boundary condition is protected against wave
interference effects, and constitutes an explicit proof that no
scaling hypothesis might exist in the extrapolation layer of
thickness $\sim l$\,. Therefore, the present work supports the
criticism of Refs.~\onlinecite{Edrei90,Lagendijk00} on the earlier
theoretical proposal \cite{Berkovits87}.

In the present work the static limit of the wave interference (weak
localization) correction namely $\delta D({\bf r}\,;
\omega\rightarrow 0)$ is explicitly calculated for the
two-dimensional semi-infinite medium with a finite width $a$ (the
bar geometry), where $\delta D({\bf r}\,; \omega\rightarrow 0)$
solely depends on the distance from the interface $z$\,. For $l\ll a
\ll le^{\pi^2\nu D_0}$ a dimensional crossover of the wave
interference correction $\delta D(z)\equiv \delta D({\bf r}\,;
\omega\rightarrow 0)$ is found. Indeed, $\delta D(z)$ displays
two-dimensional weak localization at $z\ll a$ with a logarithmic
dependence on $z$\,, while displays one-dimensional weak
localization at $a\ll z\, (\ll \xi)$\,. Furthermore, for the latter
region it is not difficult to generalize Eq.~(\ref{I1approx2}) to
higher order loop corrections which reads out as
\begin{eqnarray}
\frac{\delta D_{1{\rm D}}(z)}{D_{\rm eff}} = -\frac{z}{\xi_{1{\rm
D}}} + \sum_{n=2}^\infty\, c_n \left(\frac{z}{\xi_{1{\rm
D}}}\right)^n \,. \label{I1approx3}
\end{eqnarray}
This--at the perturbative level--formally confirms the result of
Ref.~\onlinecite{Lagendijk00} for one-dimensional geometry. Notice
that the unimportant numerical expansion coefficients $c_n$ may vary
depending on the strict/quasi- one-dimensional geometry.

For wider medium bar such that $a \gg le^{\pi^2\nu D_0}$ (e.g.,
infinite medium plane) Eq.~(\ref{I1approx}), in fact, is still
applicable except that the one-dimensional contribution vanishes.
That is,
\begin{eqnarray}
\frac{\delta D(z)}{D_0} = -\frac{1}{\pi^2\nu D_0}\,
\ln\frac{z}{l}\,. \label{I1approx4}
\end{eqnarray}
This suggests that in the medium there exists a boundary layer of
thickness $\sim l e^{\pi^2\nu D_0}$ outside which two-dimensional
strong localization occurs. Surprisingly, inside the layer the
diffusion coefficient logarithmically depends on the distance from
the interface. How to reproduce this logarithmic dependence by the
self-consistent diagrammatical method \cite{Lagendijk00} is unclear.

Finally, it should be stressed that the present field-theoretic
justification of local diffusion is perturbative. Therefore, the
validity of such a concept in the nonperturbative strong
localization region remains an important question. This problem is
far beyond the scope of this paper and will be addressed in a
forthcoming paper, especially the issue how Eq.~(\ref{I1approx3}) is
extended to the nonperturbative region: $z\gtrsim \xi$\,. It is well
known that in Faraday-active medium the one-loop weak localization
may be strongly suppressed \cite{Faraday}. Therefore, to take into
account such medium within the present field-theoretic formalism
remains another important problem. It is also interesting to
generalize the present field-theoretic formalism to include medium
gain \cite{Kroha06}. These issues are left for future work.

\acknowledgements

I am deeply grateful to A. Altland, T. Micklitz for useful
conversations, especially M. R. Zirnbauer for several fruitful
discussions. Work supported by Transregio SFB 12 of the Deutsche
Forschungsgemeinschaft.

\begin{appendix}

\section{Proof of Eq.~(\ref{theorem})}
\label{DerivationEffGreen}

Assuming that ${\bf r}\,, {\bf r}''\in {\cal V}_-$ and ${\bf
r}' \in {\cal V}_+$\,, from the Helmholtz equation (\ref{Helmholtz})
we obtain:
\begin{eqnarray}
&  &  G^R_{\Omega^2}({\bf r},{\bf r}') \{\nabla^2 +
\Omega_+^2[1+\epsilon({\bf r})]\} [g^A_{\Omega^2}({\bf r},{\bf
r}'')]^* \nonumber\\
&  =  &  \delta ({\bf r}-{\bf r}'')\, G^R_{\Omega^2}({\bf r},{\bf
r}') \label{effGF1}
\end{eqnarray}
and
\begin{eqnarray}
[g^A_{\Omega^2}({\bf r},{\bf r}'')]^*\, \{\nabla^2 +
\Omega_+^2[1+\epsilon({\bf r})]\} G^R_{\Omega^2}({\bf r},{\bf r}')
=0 \,.
\label{effGF2}
\end{eqnarray}
Subtracting Eq.~(\ref{effGF2}) from Eq.~(\ref{effGF1}) gives
\begin{eqnarray}
&& \!\!\!\!\!\!\! \nabla\cdot \{[\nabla g^{A*}_{\Omega^2}({\bf
r},{\bf r}'')] G^R_{\Omega^2}({\bf r},{\bf r}')-
 [g^{A*}_{\Omega^2}({\bf
r},{\bf r}'')]\nabla G^R_{\Omega^2}({\bf r},{\bf r}') \}
\nonumber\\
&=&\delta ({\bf r}-{\bf r}'')\, G^R_{\Omega^2}({\bf r},{\bf r}') \,,
\label{effGF3}
\end{eqnarray}
where $\nabla$ acts only on ${\bf r}$\,. Noticing that
$g^A_{\Omega^2}({\bf r},{\bf r}'')|_{{\bf r}\in C}=0$ and
$[g^A_{\Omega^2}({\bf r},{\bf r}'')]^*=g^R_{\Omega^2}({\bf r}'',{\bf
r})$\,, with ${\bf r} \in {\cal V}_-$ integrated out we find
\begin{eqnarray}
G^R_{\Omega^2}({\bf r}'',{\bf r}') = \int_C d{\bf r} \,
\partial_{{\bf n}({\bf r})} \{g^R_{\Omega^2}({\bf r}'',{\bf r})\} G^R_{\Omega^2}({\bf r},{\bf
r}') \,. \label{effGF4}
\end{eqnarray}
Here $\partial_{{\bf n}({\bf r})}$ stands for the normal derivative
at ${\bf r}$ with ${\bf n}({\bf r})$ pointing to ${\cal V}_+$\,. Taking the derivative we obtain:
\begin{eqnarray}
\partial_{{\bf n}({\bf r}'')}G^R_{\Omega^2}({\bf r}'',{\bf r}') = \int_C d{\bf r}
\, B({\bf r}'',{\bf r})\, G^R_{\Omega^2}({\bf r},{\bf r}')
\label{effGF5}
\end{eqnarray}
with $B({\bf r}'',{\bf r})$ following the definition of Eq.~(\ref{effGF6}).

Now suppose that ${\bf r}$ is shuffled to $C$ from the
${\cal V}_-$ side. Let us integrate out Eq.~(\ref{Greenfunction}) over
the line element along an infinitesimal piece of a curve passing from ${\cal V}_-$
to ${\cal V}_+$\,. In doing so we
obtain:
\begin{eqnarray}
\partial_{{\bf n}({\bf r})} G^R_{\Omega^2}({\bf r},{\bf r}')|_{{\bf r}\in C^+} -
\partial_{{\bf n}({\bf r})} G^R_{\Omega^2}({\bf r},{\bf r}')|_{{\bf r}\in C^-}=0 \,,
\label{effGF9}
\end{eqnarray}
where $C^+$ ($C^-$) stands for the curve infinitesimally closed to $C$ from the ${\cal V}_+$ (${\cal V}_-$) side.
Taking Eq.~(\ref{effGF5}) into account we may rewrite
Eq.~(\ref{Greenfunction}) as
\begin{eqnarray}
&& \left\{
 \Omega_+^2- {\hat H} \right\}\, G^{R}_{\Omega^2}({\bf r},{\bf
 r}')-\int_C d{\bf r}''
\, B({\bf r},{\bf r}'')\, G^R_{\Omega^2}({\bf r}'',{\bf r}')\nonumber\\
&& \qquad \qquad \qquad =0 \,, \quad {\bf r}\in C\,,
\label{effGF10}
\end{eqnarray}
which is supplemented by the boundary condition $\partial_{{\bf n}({\bf r})}
G^R_{\Omega^2}({\bf r},{\bf r}')|_{{\bf r}\in C}=0$\,.

$B({\bf r},{\bf r}')$ (${\bf r},{\bf r}'\in C$) consists of the real
and imaginary part. The former is small and may be absorbed into
$\Omega_+^2$ renormalizing $\epsilon({\bf r})$\,. It is thus
ignored. In contrast, the latter is important determining the
analytical structure. Taking this into account we prove
Eq.~(\ref{theorem}) for the retarded (and similarly for the
advanced) Green function.

\section{Simplification of the coupling action $F_{\rm inter}[Q]$}
\label{simplification}

For the moment let us suppress
the indices $i$ and $k_\perp$\,, and decompose
$Q$ according to
\begin{eqnarray}
&& Q=Q_\perp+Q_\parallel\,, \nonumber\\
&& [Q_\perp\,, \Lambda]=0\,, \quad \{Q_\parallel\,, \Lambda\}=0 \,.
\label{Qdecomposition}
\end{eqnarray}
Taking such decomposition into account we obtain:
\begin{eqnarray}
&& {\rm str}\ln \, (1+\alpha \Lambda Q) \label{simplification1}\\
&=& {\rm str}\ln \left(1+\alpha \Lambda Q_\parallel\right) +
{\rm str}\ln \left(1+\frac{1}{1+\alpha \Lambda Q_\parallel}\, \alpha \Lambda Q_\perp\right)\,.
\nonumber
\end{eqnarray}
Upon Taylor expanding the second logarithm only the even order terms contribute.
Thus, Eq.~(\ref{simplification1}) may be rewritten as
\begin{eqnarray}
&& {\rm str}\ln \, (1+\alpha \Lambda Q) \nonumber\\
&=& {\rm str}\ln \left(1+\alpha \Lambda Q_\parallel\right) + \frac{1}{2}\,
{\rm str}\ln \left(1+\frac{1}{1+\alpha \Lambda Q_\parallel}\, \alpha \Lambda Q_\perp\right)
\nonumber\\
&& \frac{1}{2}\,
{\rm str}\ln \left(1-\frac{1}{1+\alpha \Lambda Q_\parallel}\, \alpha \Lambda Q_\perp\right)
\nonumber\\
&=& \frac{1}{2}\,
{\rm str}\ln \left\{\left(1+\alpha \Lambda Q_\parallel\right)^2+\left(\alpha Q_\perp\right)^2\right\}\nonumber\\
&=& \frac{1}{2}\,
{\rm str}\ln \left(2+2\alpha^2 + 4\alpha\Lambda Q_\parallel\right)
\nonumber\\
&=& \frac{1}{2}\,
{\rm str}\ln \left(2-T_0 + T_0\Lambda Q_\parallel\right)\,,
\label{simplification2}
\end{eqnarray}
where in deriving the third equality we use the identity $Q_\parallel^2 + Q_\perp^2=1$\,,
and in deriving the last two equalities we use the identity ${\rm str} \ln {\bf 1}=0$\,.

Restoring the index $i$ and substituting Eq.~(\ref{simplification2})
into Eq.~(\ref{interF2}) we obtain:
\begin{eqnarray}
F_{\rm inter}[Q]
&=& -\frac{\Omega l}{4\pi}\,\sum_i
{\rm str}\ln \left[2-T_0(i) + T_0(i)\Lambda Q_{i\parallel}\right]\nonumber\\
&\approx& -\frac{\Omega l}{4\pi}\,\sum_i T_0(i)
{\rm str}\, \left(\Lambda Q_{i\parallel}\right) \nonumber\\
&=& -\frac{\Omega l}{4\pi}\,\sum_i T_0(i)
{\rm str}\, \left(\Lambda Q_i\right)\,,
\label{simplification3}
\end{eqnarray}
where in the second line we take advantage of strong coupling, i.e.,
$\Omega l\gg 1$\,, and in the third equality we use the identity
${\rm str}\, (\Lambda Q_\perp)=0$\,.

\section{The charge-conjugation symmetry of $W$}
\label{Wsymmetry}

The charge-conjugation symmetry is irrespective of fast-slow mode
separation and therefore we ignore the subscript $>(<)$\,.
Substituting the parametrization of $W$ into Eq.~(\ref{Wrelation})
we obtain:
\begin{eqnarray}
\left(\begin{array}{cc}
  0 & B \\
  {\bar B} & 0
\end{array}\right)^{ar} = \left(
                            \begin{array}{cc}
                              0 & {\bar B}^\dagger k \\
                              kB^\dagger & 0 \\
                            \end{array}
                          \right)^{ar}
\label{Wrelation1}
\end{eqnarray}
giving $kB^\dagger = {\bar B}$ and ${\bar B}^\dagger k=B$\,. The
first relation may be rewritten as
\begin{equation}
B^*k = C_0BC_0^{\rm T}\,. \label{Brelation}
\end{equation}
Substituting the second relation into Eq.~(\ref{Brelation}) we
obtain:
\begin{equation}
B^*k = C_0{\bar B}^{*{\rm T}}k C_0^{\rm T}
\label{Brelation1}
\end{equation}
giving $Bk=C_0{\bar B}^{\rm T}kC_0^{\rm T}$\,. Noticing the
relation: $C_0kC_0^{\rm T}=k$ one finds
\begin{equation}
B = C_0{\bar B}^{{\rm T}} C_0^{\rm T} = {\bar{\bar B}}
\label{Brelation2}
\end{equation}
and thus justifies the charge-conjugation symmetry of $W$\,.

\section{Preservation of Ward identity}
\label{cancelation}

Using integral by parts we transform Eq.~(\ref{F4}) into (noticing
that $\partial_\rho W({\bf r})|_{\rho=0\, {\rm or}\, a}=0$)
\begin{widetext}
\begin{equation}
F_{4}[W] = \frac{\pi\nu}{2} \int_0^{\infty}\!\!\!\! dz \!\!\int_0^a
\!\!\!\! d\rho\,\left\{ 2 D_0\, {\rm str}\, \left[\partial^2 W\,
W^3+\left(\partial W \, W\right)^2+\left(\partial W\right)^2
W^2\right] + i\omega \, {\rm str}\, W^4 \right\} \,.
\label{F4transformation}
\end{equation}
Using the contraction rule we obtain:
\begin{eqnarray}
&  &  \pi\nu D_0\, \int_0^{\infty}\!\!\!\! dz_1 \!\!\int_0^a
\!\!\!\! d\rho_1 \, \left\langle{\rm str}\, [k (1+\Lambda)(1-
\tau_3) W({\bf r}) k (1-\Lambda)(1+ \tau_3) W ({\bf r}')]\, {\rm
str}\, [\partial^2 W \,({\bf r}_1) W^3({\bf
r}_1)]\right\rangle_{F_2} \label{Ia}\\
&  =  &  I_a + \pi\nu D_0\, \int_0^{\infty}\!\!\!\! dz_1
\!\!\int_0^a \!\!\!\! d\rho_1 \, \big\langle{\rm str}\, [k
(1+\Lambda)(1- \tau_3) W({\bf r}) k (1-\Lambda)(1+ \tau_3) W ({\bf
r}')]\, {\rm str}\, \big[\partial^2 \overbrace{W \,({\bf r}_1)
W({\bf r}_1)W}\, ({\bf r}_1)W({\bf r}_1)\big] \big\rangle_{F_2} \,,
\nonumber
\end{eqnarray}
where the overbrace fixes the contraction and
\begin{eqnarray}
I_a &  =  &  \pi\nu D_0\, \int_0^{\infty}\!\!\!\! dz_1 \!\!\int_0^a
\!\!\!\! d\rho_1 \, \bigg\{\big\langle{\rm str}\, k (1+\Lambda)(1-
\tau_3) W({\bf r}) k (1-\Lambda)(1+ \tau_3) \overbrace{W ({\bf
r}')\, {\rm str}\,
\partial^2 W}\,({\bf r}_1) W^3({\bf r}_1)\big\rangle_{F_2} +
\nonumber\\
&  &  \big\langle{\rm str}\, k (1+\Lambda)(1- \tau_3) \overbrace
{W({\bf r}) k (1-\Lambda)(1+ \tau_3) W ({\bf r}')\, {\rm str}\,
\partial^2 W}\,({\bf r}_1) W^3({\bf r}_1) \big\rangle_{F_2}\bigg\}
\nonumber\\
&  =  &  D_0\, \int_0^{\infty}\!\!\!\! dz_1 \!\!\int_0^a \!\!\!\!
d\rho_1 \, \bigg\{ \partial_{{\bf r}_1}^2 {\cal D}({\bf r}', {\bf
r}_1;\omega) \big\langle{\rm str}\,  [k (1+\Lambda)(1- \tau_3)
W({\bf r}) k (1-\Lambda)(1+ \tau_3) W^3({\bf r}_1)]
\big\rangle_{F_2}+\nonumber\\
&  &  \qquad\qquad\qquad\qquad\,\,\,\,\partial_{{\bf r}_1}^2 {\cal
D}({\bf r}, {\bf r}_1;\omega)\big\langle {\rm str}\,  [k
(1+\Lambda)(1- \tau_3) W^3({\bf r}_1) k (1-\Lambda)(1+ \tau_3)
W({\bf r}')] \big\rangle_{F_2}\bigg\} \,.
\label{Iaresult}
\end{eqnarray}
Likewise, we also obtain:
\begin{eqnarray}
I_b &  \equiv  &  \frac{i\pi\nu\omega }{2}\, \int_0^{\infty}\!\!\!\!
dz_1 \!\!\int_0^a \!\!\!\! d\rho_1 \, \left\langle{\rm str}\, [k
(1+\Lambda)(1- \tau_3) W({\bf r}) k (1-\Lambda)(1+ \tau_3) W ({\bf
r}')]\, {\rm str}\,
 [W^4({\bf r}_1)] \right\rangle_{F_2}\nonumber\\
&  =  &  i\omega\, \int_0^{\infty}\!\!\!\! dz_1 \!\!\int_0^a
\!\!\!\! d\rho_1 \, \bigg\{ {\cal D}({\bf r}', {\bf r}_1;\omega)
\left\langle{\rm str}\, [k (1+\Lambda)(1- \tau_3) W({\bf r}) k
(1-\Lambda)(1+ \tau_3) W^3({\bf
r}_1)] \right\rangle_{F_2}+ \nonumber\\
&  &  \qquad\qquad\qquad\qquad{\cal D}({\bf r}, {\bf r}_1;\omega)
\left\langle{\rm str}\, [k (1+\Lambda)(1- \tau_3) W^3({\bf r}_1) k
(1-\Lambda)(1+ \tau_3) W({\bf r}')]\right\rangle_{F_2}\bigg\}\,.
 \label{Ib}
\end{eqnarray}
\end{widetext}
Noticing Eq.~(\ref{diffusion}) we find that $I_a+I_b$ exactly
cancels the first two terms of Eq.~(\ref{cooperonSL1}).

\section{Derivation of Eq.~(\ref{I1approx})}
\label{Izdiscussion}

Let us introduce the function:
$ f(z) = \sum_{n\geq 1} \, [1-e^{-2n\pi z/a}]/(n\pi)$\,.
Taking its derivative we obtain:
\begin{eqnarray}
f'(z)  =   \frac{2}{a}\, \sum_{n\geq 1} e^{-2n\pi z/a} =
\frac{2}{a}\, \frac{e^{-2\pi z/a}}{1-e^{-2\pi z/a}} \,.
\label{fderivative}
\end{eqnarray}
On the other hand, the low-energy diffusion occurs on the scale
$\sim l$\,. Over this scale the interface where $f(z)$ vanishes is
smeared. Therefore, without loss of any physics we may reformulate
the boundary condition as $f(l)=0$\,. Taking it into account and
integrating out Eq.~(\ref{fderivative}) we obtain
$ \pi f(z) = \ln \{(1-e^{-2\pi z/a})/(1-e^{-2\pi l/a})\} $
for $z\gtrsim l$\,, which gives $f(z) \approx \pi^{-1}\ln (z/l)$ for
$l\lesssim z\ll a$ justifying Eq.~(\ref{I1approx}).

\end{appendix}


\end{document}